\documentstyle[prl,aps,]{revtex}
\begin{document}
\title
{Reply to the Comment on  ``Bound States in the 
One-dimensional Hubbard Model''}
\author{Daniel Braak$^{1,2}$ and Natan Andrei$^{1}$}
\address{$^1$ Department of Physics and Astronomy, Rutgers University,
Piscataway, NJ 08855}
\address{$^2$ NEC Research Institute, 4 Independence Way, Princeton,
NJ 08540}
\date{\today}
\maketitle
\begin{abstract}
We reply to the comment (cond-mat/9806125) by E{\ss}ler, G\"ohmann and Korepin,
and show that their points are unfounded.
\end{abstract}
\def\p{\phi}
\def\P{\Phi}
\def\e{\eta}
\def\ep{\epsilon}
\def\vep{\varepsilon}
\def\ps{\psi}
\def\a{\alpha}
\def\ab{{\tilde{a}}}
\def\b{\beta}
\def\y{{\tilde{y}}}
\def\k{\kappa}
\def\psd{\ps^{\dagger}}
\def\psdt{{\tilde{\ps}}^{\dagger}}
\def\pst{\tilde{\ps}}
\def\di{d^{\dagger}}
\def\d{\delta}
\def\g{\gamma}
\def\be{\begin{equation}}
\def\ee{\end{equation}}
\def\bdm{\begin{displaymath}}
\def\edm{\end{displaymath}}
\def\s{\sigma}
\def\bea{\begin{eqnarray}}
\def\eea{\end{eqnarray}}
\def\bear{\begin{array}}
\def\eear{\end{array}}
\def\nn{\nonumber}
\def\x{\chi}
\def\ra{\rightarrow}
\def\r{\rho}
\def\s{\sigma}
\def\vs{\vec{\s}}
\def\S{\bar{S}}
\def\Sr{\tilde{S}}
\def\z{\zeta}
\def\Z{\bar{Z}}
\def\Zr{\tilde{Z}}
\def\ta{\tilde{A}}
\def\t{\tau}
\def\th{\theta}
\def\Th{\Theta}
\def\l{\lambda}
\def\L{\Lambda}
\def\K{{\cal K}}
\def\Li{{\cal L}}
\def\E{{\cal E}}
\def\ki{\frac{k_i}{\L}}
\def\kj{\frac{k_j}{\L}}
\def\ua{\uparrow}
\def\da{\downarrow}
\def\rd{\textrm{d}}
\def\id{{\mathbf{1}}}
\def\ph{\phantom}
\def\O{{\cal{O}}}
\def\o{\omega}
\def\H{\tilde{H}}
\def\arsinh{{\textrm{arsinh}}}
\def\cotanh{{\textrm{cotanh}}}
\def\arcsin{{\textrm{arcsin}}}
\def\sech{{\textrm{sech}}}
\def\vu{\frac{4}{u}}
\def\uv{\frac{u}{4}}
\def\uz{\frac{u}{2}}
\def\cz{\frac{c}{2}}
\def\dx{\Delta\x}
\def\ecl{e^{-\frac{|c|}{2}L}}
\def\eclp{e^{\frac{|c|}{2}L}}
\def\eul{e^{-\frac{|u|}{4}L}}
\def\eulp{e^{\frac{|u|}{4}L}}
\def\ekl{e^{-\xi L}}
\def\eklp{e^{\xi L}}
\def\ekld{e^{-2\xi L}}
\def\ln{{\textrm{ln}}}
\def\G{{\hat{G}}}
\def\ran{\rangle}

\bigskip

We  thank the authors of the comment \cite{eskor} 
for correcting some typos in our paper \cite{bran}. They have missed however
 the main point. Bound states 
cannot be
introduced into the Lieb-Wu equations (which have been derived for real $k$ with
boundary conditions corresponding to a ring of finite length $L$) simply
 by inserting
$k$-$\Lambda$-strings. Bound states correspond to {\it poles} of the S-matrix,
a fact that guarantees that the wave-functions with complex $k$ do not blow
up in the $L \rightarrow \infty$ limit. Our approach provides a correct and general procedure
for incorporating complex momenta. Furthermore, there is no need to appeal  to
 a "string
hypothesis".

We proceed to  respond  in turn to the points  raised in the comment. 

(I) Our Bethe equations (eq.(20)-(22) in \cite{bran}) are not new
but ``coincide'' with Takahashi's equations (eq.(2.11a-c) in
\cite{taka})
for finite system size $L$.

This is incorrect: The equations are  different for $L$ {\it finite}.
Takahashi has dropped terms of order
$\eul$ in his equations (2.11a-c). 
He started from the Lieb-Wu equations \cite{liebwu} (eq.(2,3) in \cite{bran})
for a finite system with
periodic boundary conditions (PBC) and used the
"string hypothesis" for both the  $k$-$\Lambda$-strings and the
$\Lambda$-strings. Keeping the neglected terms explicitly one
arrives first at eq.(5)-(8) in our paper (these are simplified as they
contain only 1-complexes and no $\L$-strings). Eliminating from the
set (5)-(8) the variables $\varphi_l$ and dropping the $\vep$-terms
leads to (2.11a-c) in \cite{taka}. This procedure is 
incorrect because it assumes the
consistency of (5)-(8) for finite $L$. Our approach is different: We do not proceed from (2,3)
but derive the analogue to the Lieb-Wu equations (the BAC) for composite
boundary conditions (CBC), which allow us to include the bound states
ab initio. For an $m$-complex these boundary conditions read 
\[
F(\{x_j\},x^a_1,\ldots ,x^a_{2m})=F(\{x_j\},x^a_1+L,\ldots,x^a_{2m}+L)
\]
where $x^a_1\ldots x^a_{2m}$ denote the positions of the particles
forming
the $m$-complex, and $\{x_j\}$ the positions of the
 $N-2m$ others. 
These boundary conditions coincide with PBC for unbound particles.
 Our equations (20-22) (BAC equations)
are {\it exact} for finite $L$ and CBC. Moreover,
 the reason why they do not contain
the $\L$-strings (which the authors of \cite{eskor} suspect to be
``hidden'' in our notation) is  that these (spin-) strings follow
 from a hypothesis, which is known not to be always true  
\cite{low,woyn3}.

(II)\ The authors repeat a calculation first performed in 
\cite{woyn1} to show that in case of a single bound state
excitation the relation $2\L'=\sin k^h_1+\sin k^h_2$ is satisfied
at half-filling. This statement has no physical meaning as is
explained in column 2 on page 4 in \cite{bran}: The bound state
excitation is an independent excitation only
away from half-filling.

(III)\ The authors believe that we 
wanted to keep eq.(5),(6) and (8), while objecting to (7). This guess is
false, as (5) and (8) contain the spin-rapidities $\L_l$, whose presence
causes the overdetermination of the set (5)-(8). Their
derivation of the equivalence of the set (5)-(8) with the set
(5),(6),(8),(B8) in appendix B of \cite{eskor} does not prove the consistency
of one of these sets, which follow from
the $k$-$\L$-string hypothesis - the fact that (B8) is always true
for translational invariant systems with overall $L$-periodicity 
is
obviously insufficient. In any case, point (III) in \cite{eskor}
has nothing to do with our argument for the redundancy
of (7) in the infinite volume limit, as this argument does not relate
to periodicity but to the equations (3) determining the conditions on
the
eigenvectors of
the transfermatrix.

 (IV) It is interesting to note that 
the reasoning  of \cite{eskor} (if correct) would lead
to the following alternatives:

1)\ Either the Hilbert spaces 
for PBC respectively CBC have the same dimension
as we claim
and point (IV) is empty, or

2)\ the dimension for CBC is larger, which entails an error in
\cite{kores}, because there the number of states with CBC is counted
(remember that the BAC equations are not based on a
hypothesis),
which is, according to \cite{kores}, $4^L$.

We wish to thank F. E{\ss}ler and V. Korepin for interesting correspondance.

\end{document}